\documentclass[12pt]{article}

\usepackage{times}

\newenvironment{sciabstract}{%
\begin{quote} \bf}
{\end{quote}}

\newcounter{lastnote}

\title{Magnetic Resonant Mode in the Single-Layer High Temperature Superconductor
Tl$_2$Ba$_2$Cu$_{6+\delta}$}

\author
{H. He,$^{1}$ P. Bourges,$^{2}$ Y. Sidis,$^{2}$ C. Ulrich,$^{1}$
L.P. Regnault,$^{3}$\\
S. Pailh\`{e}s,$^{2}$ N.S. Berzigiarova,$^{4}$ N.N. Kolesnikov,$^{4}$ B.~Keimer$^{1\ast}$\\
\\
\normalsize{$^{1}$Max-Planck-Institut f\"{u}r
Festk\"{o}rperforschung, 70569 Stuttgart, Germany.}\\
\normalsize{$^{2}$Laboratoire L\'{e}on Brillouin, CEA-CNRS, CE
Saclay, 91191 Gif sur Yvette, France.}\\
\normalsize{$^{3}$CEA Grenoble, D\'{e}partement de Recherche
Fondamentale sur la Mati\`{e}re Condens\'{e}e,}\\
\normalsize{38054 Grenoble cedex 9, France.}\\
\normalsize{$^{4}$Institute of Solid State Physics, Russian
Academy of Science,}\\
\normalsize{Chernogolovka, 142432 Russia.}\\
\\
\normalsize{$^\ast$To whom correspondence should be addressed;
E-mail:  b.keimer@fkf.mpg.de.}
}

\date{}

\begin{document}

\baselineskip24pt

\maketitle

\newpage

\begin{sciabstract}

An unusual spin excitation mode observed by neutron scattering has
inspired numerous theoretical studies of the interplay between
charged quasiparticles and collective spin excitations in the
copper oxide high temperature superconductors. The mode has thus
far only been observed in materials with crystal structures
consisting of copper oxide bilayers, and it is notably absent in
the single-layer compound La$_{2-x}$Sr$_{x}$CuO$_{4+\delta }$.
Neutron scattering data now show that the mode is present in
Tl$_2$Ba$_2$Cu$_{6+\delta}$, a single-layer compound with T$_c$
$\sim$ 90 K, thus demonstrating that it is a generic feature of
the copper oxide superconductors, independent of the layer
sequence. This restricts the theoretical models for the origin of
the resonant mode and its role in the mechanism of high
temperature superconductivity.

\end{sciabstract}

\newpage

Electronic conduction in the copper oxide high temperature
superconductors takes place predominantly in structural units of
chemical composition CuO$_{2}$, in which copper and oxygen atoms
form an approximately square planar arrangement. Most theoretical
models of high temperature superconductivity are therefore based
on a two-dimensional square lattice. In real materials, however,
deviations from this simple situation are nearly always present.
For instance, buckling distortions of the CuO$_{2}$ layers that
are found in many copper oxides are thought to have a significant
influence on the electronic structure and on the superconducting
transition temperature $\rm T_{c}$. Interlayer interactions in
materials with closely spaced CuO$_{2}$ layers (forming bi- or
trilayer units) or additional copper oxide chains in the crystal
structure present further complications whose influence on the
superconducting
properties remains a subject of debate. Experiments on Tl$_{2}$Ba$_{2}$CuO$%
_{6+\delta }$, a material with unbuckled, widely spaced CuO$_{2}$
layers and a maximum $\rm T_{c}$ around 90 K, have therefore
played a pivotal role in resolving some issues central to our
understanding of these materials ({\it 1, 2\/}). We report
inelastic neutron scattering measurements of
Tl$_{2}$Ba$_{2}$CuO$_{6+\delta } $ near optimum doping  ($\rm
T_{c} \sim 90$ K) that provide evidence of a sharp magnetic
resonant mode below $\rm T_{c}$.

A resonant spin excitation of this kind has been extensively
characterized by neutron scattering in the bilayer copper oxide
YBa$_{2}$Cu$_{3}$O$_{6+\delta }$ ({\it 3--5\/})
and was recently also observed in Bi$_{2}$Sr$_{2}$%
CaCu$_{2}$O$_{8+\delta }$, another bilayer compound ({\it 6,
7\/}). At all doping levels, strong line shape anomalies of this
collective spin excitation below T${\rm _{c}}$ bear witness to a
substantial interaction with charged quasiparticles. Conversely,
anomalies in the quasiparticle spectra observed by photoemission
({\it 8, 9\/}), optical conductivity ({\it 10--13\/}), tunneling
({\it 13, 14\/}), and Raman scattering ({\it 15\/}) techniques
have been interpreted as evidence of coupling to the neutron mode.
In the copper oxides (as well as some heavy fermion materials,
where similar observation have been made ({\it 16\/}) the
correspondence between anomalous features in the spectra of spin
and charge excitations has stimulated spin fluctuation based
pairing scenarios ({\it 9, 11, 16\/}) which are, however, still
controversial ({\it 17--19\/}). Other theories of high temperature
superconductivity also rely on the resonant mode ({\it 20\/}). It
is therefore of great interest to establish whether the resonant
spin excitation is a general feature of the various
crystallographically distinct families of superconducting copper
oxides. The failure to detect such an excitation in the
single-layer compound La$_{2-x}$Sr$_{x}$CuO$_{4+\delta }$, despite
much experimental effort ({\it 21\/}), has hampered a unified
phenomenology of the copper oxides, and the prospect that the mode
could be a spectral feature specific to bilayer materials has cast
a cloud over models in which spin excitations play a central role.
The result of our experiment on single-layer Tl$_{2}$Ba$%
_{2}$CuO$_{6+\delta }$\ (whose T${\rm _{c}} \sim 90$ K is closely
similar
to that of optimally doped YBa$_{2}$Cu$_{3}$O$_{6+\delta }$ and Bi$_{2}$Sr$%
_{2}$CaCu$_{2}$O$_{8+\delta }$) implies that strong magnetic
interactions between closely spaced CuO$_{2}$ layers are not
required for the formation of the resonant mode. The different
form of the spin excitation spectrum of
La$_{2-x}$Sr$_{x}$CuO$_{4+\delta }$ may be due to the proximity of
a competing instability ({\it 22\/}) that could also be
responsible for the anomalously low T${\rm _{c}}\leq $ 40K.

Inelastic magnetic neutron scattering measurements on conventional
triple axis or time-of-flight spectrometers require high quality
copper oxide single crystals with volumes of at least 1 cm$^{3}$,
which are only
available for the La$_{2-x}$Sr$_{x}$CuO$_{4+\delta }$ and YBa$_{2}$Cu$_{3}$O$%
_{6+\delta }$ families. Due to advances in focusing techniques
that improve the neutron beam delivery onto small samples,
experiments on single crystals of
Bi$_{2}$Sr$_{2}$CaCu$_{2}$O$_{8+\delta }$ with an order of
magnitude smaller volumes have recently become feasible ({\it 6,
7\/}). Unfortunately, the crystal growth of the Tl- and Hg-based
copper oxide
superconductors (which include the only known single-layer compounds with T$%
{\rm _{c}}$ of order 90 K) has suffered from severe technical
difficulties, chief among them the toxicity of some of the
constituents. With typical crystal volumes well below 1 mm$^{3}$,
inelastic neutron scattering on these compounds has thus far
appeared hopeless. We have overcome these difficulties by
synthesizing about 300 relatively large (0.5--3 mm$^{3}$) single
crystals of Tl$_{2}$Ba$_{2}$CuO$_{6+\delta }$ through a CuO-rich
flux technique ({\it 23\/}), and co-aligning them in a mosaic of
total volume 0.11 cm$^{3}$\ using Laue x-ray diffraction (Fig.
1A). The crystallographic axes of the individual crystals in the
array were aligned with an accuracy of about 1.5$^{\circ }$ (Fig.
1C). Before alignment, the magnetic susceptibilities of all
crystals were measured as a function of temperature (typical data
are shown in Fig. 1B). The mean T${\rm _{c}}$ (onset)\ of \ the
individual crystals used in the array was 92.5 K, with a standard
deviation of 2 K.

Compared to
previous neutron scattering experiments on much larger crystals of Bi$%
_{2}$Sr$_{2}$CaCu$_{2}$O$_{8+\delta }$, the background from the Al
plates and the adhesive holding the multicrystal array (Fig. 1A)
was significantly enhanced, while the signal amplitude is
substantially smaller due to the single-layer structure (see
below). In order to compensate for this reduction in
signal-to-background ratio, counting times of up to 5 hours per
data point had to be used. Of particular interest was the energy
range
around 40 meV, where the resonant mode was observed in both YBa$_{2}$Cu$_{3}$%
O$_{6+\delta }$ and Bi$_{2}$Sr$_{2}$CaCu$_{2}$O$_{8+\delta }$ near
optimum doping. In these two materials, the mode manifests itself
in a sharply enhanced
intensity around the in-plane wave vector ${\bf Q}_{\rm AF}=(\frac{1}{2},\frac{1}{%
2})$ that is present only below T${\rm _{c}}$; when the doping is
reduced, magnetic intensity is observed also above T${\rm _{c}}$.
(The components of the wave vector are denoted as ${\bf
Q}=(H,K,L)$ and given in units of the reciprocal lattice vectors
$a^{\ast }=b^{\ast }=1.62$ \AA $^{-1}$ and $c^{\ast }=0.27$ \AA
$^{-1}$ of the tetragonal structure.) As a function of the wave
vector component {\it L }perpendicular to the CuO$_{2}$ layers,
one observes an intensity modulation due to strong interlayer
interactions within a bilayer unit ({\it 3--7\/}). As the
Tl$_{2}$Ba$_{2}$CuO$_{6+\delta }$ are much further apart and hence
much more weakly interacting, it is expected that this modulation
is absent, and that the magnetic intensity is independent of {\it
L} except for a slow decrease with increasing {\it L} due to the
magnetic form factor. Note that the absence of the bilayer
modulation goes along with a factor-of-two reduction in signal
amplitude, which is further compounded by the lower density of
CuO$_2$ planes in the Tl$_{2}$Ba$_{2}$CuO$_{6+\delta }$ structure.

The constant-energy scans (Fig. 2) indeed exhibit this
characteristic signature, albeit at an energy of 47 meV that is
somewhat larger than 40 meV, the mode energy in the optimally
doped bilayer compounds; constant-energy scans around 40 meV were
featureless (Fig. 2D). These data were taken on two different
spectrometers with different final energies, with identical
results. Above T${\rm _{c}}$, the scans show a
featureless background that gradually decreases in an energy- and {\bf Q}%
-independent fashion as the temperature is lowered. In the
superconducting state, a sharp peak centered at {\bf Q}$_{\rm AF}$
appears on top of this background. As expected for magnetic
scattering that is uncorrelated from
layer to layer, the peak intensities measured at two inequivalent {\it L}%
-positions ($L = 10.7$ and $12.25$, respectively) are identical
within the errors. The identification of this peak with the
magnetic resonant mode is further supported by comparing
constant-{\bf Q} scans at ${\bf Q}={\bf Q}_{\rm AF}$ above and
below T${\rm _{c}}$. The difference signal shows a peak that is
resolution limited in energy (Fig. 3), in agreement with
observations in optimally doped YBa$_{2}$Cu$_{3}$O$_{6+\delta }$.

Constant-energy scans taken at different temperatures were fitted
to Gaussian profiles (Fig. 2). The best overall fit was provided
by an intrinsic {\bf Q}-width of 0.23 $\pm$ 0.05 \AA$^{-1}$ (full
width at half maximum). The data were subsequently refitted with
the {\bf Q}-width kept fixed in order to reduce scatter, and the
peak amplitude thus extracted as a function of temperature (Fig.
4). There is some indication of a broad peak centered around {\bf
Q}$_{\rm AF}$ even above T${\rm _{c}}$, as observed in underdoped
YBa$_{2}$Cu$_{3}$O$_{6+\delta }$ ({\it 3-5\/}), but this is well
within the experimental error. Based on a calibration measurement
of acoustic phonons (Fig. 1D), the spectral weight of the mode
(that is, the energy-integrated intensity) at {\bf Q}$_{\rm AF}$
is determined as $0.7 \pm 0.25$ $\mu_B^2$, which corresponds to
$0.02 \pm 0.007$ $\mu_B^2$ when averaged over the entire Brillouin
zone. Both the {\bf Q}-width and the absolute spectral weight per
$\rm CuO_2$ plane of the peak observed in
Tl$_{2}$Ba$_{2}$CuO$_{6+\delta }$
are identical within the errors to those of the resonant mode measured in YBa$%
_{2}$Cu$_{3}$O$_{7}$ ({\it 4, 24\/}), as are its dependence on
energy and temperature. Although for
Tl$_{2}$Ba$_{2}$CuO$_{6+\delta }$ the experimental information is
much more limited at this time than for the latter material (large
crystals of which have been studied for more than a decade), the
cumulation of independent matching features allows us to infer
with confidence a common origin of both phenomena. Other possible
explanations of the observations in Tl$_{2}$Ba$_{2}$CuO$_{6+\delta
}$ (such as phonon anomalies) would require an implausible
coincidence, especially since above ${\rm T_c}$ both the raw
constant-energy scans (Fig. 2B) and the raw constant-{\bf Q} scans
around 47 meV (Fig. 3A) are featureless.

It is interesting to note that the resonant modes of both YBa$_{2}$%
Cu$_{3}$O$_{7}$ and Tl$_{2}$Ba$_{2}$CuO$_{6+\delta }$ are
resolution limited
in energy whereas a significant broadening is found in Bi$_{2}$Sr$_{2}$CaCu$%
_{2}$O$_{8+\delta }$ ({\it 6, 7\/}). As even small concentrations
of impurities systematically introduced into
YBa$_{2}$Cu$_{3}$O$_{7}$ broaden the mode considerably ({\it 24,
25\/}), the broadening in Bi$_{2}$Sr$_{2}$CaCu$_{2}$O$_{8+\delta
}$ was attributed to disorder. The
sharpness of the mode observed in optimally doped Tl$_{2}$Ba$_{2}$CuO$%
_{6+\delta }$ appears to imply that disorder effects are minimal
in this material, a surprising finding in view of the fact that Tl$_{2}$Ba$_{2}$CuO$%
_{6+\delta }$ (in contrast to YBa$_{2}$Cu$_{3}$O$_{7}$) is
nonstoichiometric. It also means that the model of Refs. ({\it 11,
12\/}), though qualitatively correct in predicting the presence of
a resonant spin excitation in Tl$_{2}$Ba$_{2}$CuO$_{6+\delta }$,
needs to be re-evaluated in view of the fact that the
experimentally determined peak width is much smaller than
predicted on the basis of the optical conductivity
data. The mode energy found in Tl$%
_{2}$Ba$_{2}$CuO$_{6+\delta }$ notably exceeds the ones in both YBa$_{2}$Cu$%
_{3}$O$_{7}$ and Bi$_{2}$Sr$_{2}$CaCu$_{2}$O$_{8+\delta }$. This
is in agreement with models according to which the resonant mode
is a collective excitation pulled below the quasiparticle pair
production continuum by exchange interactions ({\it 8, 9,
26--29\/}). The lower mode energy in the bilayer materials could
thus be a consequence of the strong interactions between the two
CuO$_{2}$ layers that form a bilayer unit ({\it 26--29}), assuming
that the energy gap and hence the pair production threshold are
identical in materials with identical T${\rm _{c}}$ as indicated
by electronic Raman scattering results ({\it 30\/}).

The most important implication of the findings reported here
regards the unified phenomenological picture recently developed
for spin and charge spectroscopies of the copper oxides ({\it
8--14\/}). The spectral anomalies that have been interpreted as
evidence of coupling to the collective spin excitation are also
present in optical conductivity ({\it 31\/}) and tunneling ({\it
32\/}) data on single-layer Tl$_{2}$Ba$_{2}$CuO$_{6+\delta }$, and
equally pronounced as in analogous data on bilayer materials. If
the mode had turned out to be absent
(or its spectral weight substantially diminished) in Tl$_{2}$Ba$_{2}$CuO$%
_{6+\delta }$, this general approach would have become untenable.
It is now time to further refine the description of the coupled
spin and charge excitations in the cuprates, and to fully evaluate
its implications for the mechanism of high temperature
superconductivity.

\newpage

\noindent {\bf References and Notes}

\begin{enumerate}

\item C.C. Tsuei {\it et al.}, {\it Nature} {\bf 387}, 481 (1997).

\item A.A. Tsvetkov {\it et al.}, {\it Nature} {\bf 395}, 360 (1998).

\item P. Bourges  in {\it The Gap Symmetry and Fluctuations in High
Temperature Superconductors} (eds. J. Bok, G. Deutscher, D. Pavuna
and S.A. Wolf) 349 (Plenum Press, New York, 1998), and references
therein.

\item H.F. Fong {\it et al}., {\it Phys.\ Rev. B} {\bf 61},
14773 (2000), and references therein.

\item P. Dai, H.A. Mook, R.D. Hunt, F. Dogan,  {\it Phys. Rev. B}
{\bf 63}, 054525 (2001), and references therein.

\item H.F. Fong {\it et al.}, {\it Nature} {\bf 398}, 588 (1999).

\item H. He {\it et al.}, {\it Phys. Rev. Lett.} {\bf 86},
1610 (2001).

\item M. Eschrig, M.R. Norman, {\it Phys. Rev. Lett}. {\bf 85},
3261 (2000), and references therein.

\item A. Abanov, A.V. Chubukov, {\it Phys.\
Rev. Lett.} {\bf 83}, 1652 (1999).

\item D. Munzar, C. Bernhard, M. Cardona, {\it Physica C} {\bf 318},
547 (1999).

\item J.P. Carbotte, E. Schachinger, D.N. Basov, {\it Nature}
{\bf 401}, 354 (1999).

\item E. Schachinger, J.P. Carbotte, {\it Phys. Rev.
B} {\bf 62}, 9054 (2000).

\item D. Manske, I. Eremin, K.H. Bennemann, {\it Phys. Rev. B}
{\bf 63}, 054517 (2001).

\item J.F. Zasadzinski {\it et al., } {\it Phys. Rev. Lett.} {\bf
87}, 067005 (2001).

\item F. Venturini, U. Michelucci, T.P. Devereaux, A.P. Kampf,
{\it Phys.\ Rev. B} {\bf 62}, 15204 (2000).

\item N.K. Sato {\it et al.}, {\it Nature} {\bf 410}, 340 (2001).

\item R. Zeyher, A. Greco, {\it Phys. Rev. B} {\bf 64}, 140510 (2001).

\item A. Lanzara {\it et al.}, {\it Nature} {\bf 412}, 510
(2001).

\item H.-Y. Kee, S.A. Kivelson, G. Aeppli, preprint cond-mat/0110478 (unpublished).

\item S.C. Zhang, {\it Science} {\bf 275}, 1089 (1997).

\item M.A. Kastner, R.J. Birgeneau, G. Shirane, Y. Endoh, {\it Rev. Mod. Phys.}
{\bf 70}, 897 (1998), and references therein.

\item J.M. Tranquada, B.J. Sternlieb, J.D. Axe, Y. Nakamura, S. Uchida,
{\it Nature} {\bf 375}, 561 (1995).

\item N.N. Kolesnikov {\it et al.}, {\it Physica C} {\bf 242},
385 (1995).

\item H.F. Fong {\it et al.}, {\it
Phys. Rev. Lett.} {\bf 82}, 1939 (1999).

\item Y. Sidis {\it et al.}, { \it
Phys. Rev. Lett.} {\bf 84}, 5900 (2000).

\item I.I. Mazin, V.M. Yakovenko, {\it
Phys. Rev. Lett.} {\bf 75}, 4134 (1995).

\item D.Z. Liu, Y. Zha, K. Levin, {\it Phys. Rev. Lett.} {\bf 75}, 4130
(1995).

\item A.J. Millis, H. Monien, {\it Phys. Rev. B} {\bf 54}, 16172 (1996).

\item F. Onufrieva, P. Pfeuty, preprint
cond-mat/9903097 (unpublished).

\item L.V. Gasparov, P. Lemmens,  M.  Brinkmann, N.N.
Kolesnikov, G. G\"untherodt, {\it Phys. Rev. B} {\bf 55}, 1223
(2000).

\item A.V. Puchkov, P. Fournier, T. Timusk, N.N. Kolesnikov,
{\it Phys. Rev. Lett.} {\bf 77}, 1853 (1996).

\item L. Ozyuzer {\it et al.}, {\it Physica C} {\bf 320}, 9 (1999).

\item We would like to thank Bernard
Hennion for his support during the LLB experiment and Patrick
Baroni for technical assistance. The work at the Institute for
Solid State Physics was supported in part by the Russian
Foundation for Basic Research. The work at the MPI-FKF was
supported in part by the German Federal Ministry of Research and
Technology (BMFT).
\end{enumerate}

\newpage

\noindent {\bf Figure Captions}

\noindent {\bf Fig. 1.} (A) Photograph of the array of co-oriented
Tl$_{2}$Ba$_{2}$CuO$_{6+\delta }$ single crystals. The crystals
are glued onto Al plates only two of which are shown for clarity.
(B) Typical diamagnetic shielding curves measured by SQUID
magnetometry on individual crystals. (C) Rocking curve of the
entire multicrystal array around the (1,1,0) Bragg reflection. The
line is a Gaussian with a full width at half maximum of
1.5$^\circ$.  (D) Constant-energy scan along ${\bf Q}=(2,2,L)$ at
excitation energy 3.1 meV and temperature 100 K, showing two
counterpropagating transverse acoustic phonons. The wave vector
{\bf Q} is given in reciprocal lattice units (r.l.u.). The data
were taken at the 2T1 triple axis spectrometer at the Orph\'{e}e
research reactor in Saclay, France, with a pyrolytic graphite (PG)
monochromator and a PG analyser set at the (002) reflection, and a
fixed final energy of 14.7 meV. Three PG filters were inserted
into the neutron beam in order to eliminate higher order
contamination. The asymmetry of the profile is a resolution
effect, and the solid line shows the result of a numerical
convolution of a standard acoustic phonon dispersion with the
spectrometer resolution.

\noindent {\bf Fig. 2.}  (A,B) Constant-energy scans along ${\bf
Q}=(H,H,L)$ with $L=10.7$ at an energy of 47.5 meV, taken on 2T
under the same conditions as the data of Fig. 1D. (C,D)
Constant-energy scans at 47 and 40 meV, respectively, taken on the
IN22 spectrometer at the Institut Laue-Langevin in Grenoble,
France, with a PG (002) monochromator and analyser and fixed final
energy 30.5 meV. The wave vector {\bf Q} is given in reciprocal
lattice units (r.l.u.). The background revealed by scans above
$\rm T_c$ (panel B) remained constant upon lowering the
temperature except for a {\bf Q}-independent scale factor. The
lines in the upper panels are the results of fits as described in
the text.

\noindent {\bf Fig. 3.} Constant-{\bf Q} scans at ${\bf
Q}=(0.5,0.5,12.25)$ taken under the same conditions as the data of
Fig. 1D. (A) Raw data at $\rm T =99 K$ ($\rm > T_c$) and $\rm
T=27$ K ($\rm < T_c$). (B) Difference between the two scans of
panel A. The line is a Gaussian profile whose width equals the
experimental energy resolution.

\noindent {\bf Fig. 4.} Temperature dependence of the magnetic
intensity extracted from fits to constant-energy scans at energy
47.5 meV around the in-plane wave vector ${\bf Q_{AF}}=(1/2,1/2)$
(Fig. 2). The data were taken on two different spectrometers (2T
and IN22) at different components of the out-of-plane wave vector
$L$.

\end{document}